\documentstyle[epsf]{article}

\title{The ground state properties 
       \protect\\
       of the spin--$\frac{1}{2}$ transverse Ising chain
       \protect\\
       with periodically varying bonds and fields}

\author{Oleg Derzhko
\footnote{On leave of absence from
Institute for Condensed Matter Physics,
1 Svientsitskii Str., L'viv--11, 79011, Ukraine
and
Chair of Theoretical Physics, 
Ivan Franko National University in L'viv,
12 Drahomanov Str., L'viv--5, 79005, Ukraine}\\
\small {Max--Planck--Institut f\"{u}r Physik komplexer Systeme,}\\
\small {N\"{o}thnitzer Str. 38, 01187 Dresden, Germany}}

\date{\today}

\begin{document}

\renewcommand\baselinestretch {1.5}
\large\normalsize

\maketitle

\begin{abstract}

Using continued fractions we study 
the ground state properties
of the spin--$\frac{1}{2}$ Ising chain 
in a transverse field 
with periodically varying interaction strengths and external fields.
We consider in detail the chain 
having the period of modulation of interactions equals 2
and compare the results obtained  
with those corresponding to 
the spin--$\frac{1}{2}$ isotropic $XY$ chain
in a transverse field.
In contrast to the behaviour of the transverse $XY$ chain,
the transverse Ising chain 
does not exhibit 
a step--like  
magnetization vs. field 
dependence  
caused by the alternation of bonds, 
its susceptibility exhibits a logarithmic singularity
at the field determined by interaction strengths,
and 
it is stable with respect to spin--Peierls dimerization.
\end{abstract}

\vspace{1cm}


\vspace{1cm}


\vspace{1mm}

\noindent
{\bf Postal address for correspondence:}\\
Dr. Oleg Derzhko\\
Institute for Condensed Matter Physics,\\
1 Svientsitskii Str., L'viv--11, 79011, Ukraine\\
Tel: (0322) 70 74 39\\
Fax: (0322) 76 19 78\\
E-mail: derzhko@icmp.lviv.ua

\clearpage

\renewcommand\baselinestretch {1.65}
\large\normalsize

Spin--$\frac{1}{2}$ $XY$ chains provide an excellent ground
for a rigorous study of different properties
of the low--dimensional quantum magnetic systems
due to the fact that 
with the help of the Jordan--Wigner transformation
such spin models can be mapped
onto noninteracting spinless fermions
and as a result
many statistical mechanics calculations can be performed exactly.$^1$
In what follows we shall consider
the periodic alternating  
$XY$ chain in a transverse field 
with {\em {extremely anisotropic}} spin coupling
(in short called here the transverse Ising chain)
in order to study the generic effects
induced by regular alternation
of the nearest neighbour interaction strengths 
and the external fields. 
The periodic alternating chain can be viewed
as a chain having several sublattices.
There are few papers dealing with the statistical mechanics properties
of $XY$ chains on two sublattices.$^{2-5}$
We want to emphasize that in the above mentioned limit
of the transverse Ising chain
the calculation of the thermodynamic quantities  
can be performed in a rather general manner
covering all  
chains with an arbitrary period of alternation.
The present paper can be viewed
as a nontrivial extension of the recent study 
of the thermodynamics  
of the spin--$\frac{1}{2}$ {\em {isotropic}} $XY$ chain 
in a transverse field 
(called here the transverse $XY$ chain)
with regularly alternating fields and bonds.$^6$
In particular,  
we compare below
the zero temperature dependences
transverse magnetization vs. transverse field
and static transverse susceptibility vs. transverse field  
for the regularly alternating transverse Ising and $XY$ chains.
On the other hand,
it is known that 
the transverse $XY$ chain 
is a simple system
exhibiting a spin--Peierls instability.$^{7-10}$ 
An analysis of the ground state energy 
of the periodic alternating transverse Ising chain 
of period 2
allows one 
to examine a spin--Peierls instability 
of that chain 
with respect to 
dimerization. 
We find that
the transverse Ising chain 
is stable with respect to dimerization
which demonstrates a role of the anisotropy in spin coupling
for a spin--Peierls instability.

We consider the $N\to\infty$ spins $\frac{1}{2}$ on a ring
with the Hamiltonian of the nonuniform transverse Ising model
\begin{eqnarray}
\label{001}
H=\sum_{n=1}^{N}\Omega_ns_n^z
+2\sum_{n=1}^{N}I_n s^x_ns^x_{n+1},
\;\;\;
s_{N+1}^{\alpha}=s_1^{\alpha}.
\end{eqnarray}
We assume {\em regular} nonuniformity in (\ref{001}),
i.e. the transverse field
$\Omega_{n}$ at the site $n$
as well as the exchange coupling $I_n$
between the neighbouring sites $n$ and $n+1$
vary regularly from site to site with period $p$,
i.e.
we have a sequence of parameters
$\Omega_{1}I_1\Omega_{2}I_2\ldots\Omega_{p}I_p
\Omega_{1}I_1\Omega_{2}I_2\ldots\Omega_{p}I_p\ldots\;$.

Our goal is to examine the thermodynamic quantities of spin model 
(\ref{001}). 
In order to do this, we shall first express the Hamiltonian 
in terms of fermion operators. 
This can be done in the usual way 
by applying the Jordan--Wigner transformation.$^1$ 
As a result 
one gets a model of spinless fermions on a ring
described by the Hamiltonian
\begin{eqnarray}
\label{002}
H=\sum_{n=1}^N\Omega_n\left(c_n^+c_n-\frac{1}{2}\right)
+\frac{1}{2}\sum_{n=1}^NI_n
\left(c_n^+c_{n+1}^++c_n^+c_{n+1}-c_nc_{n+1}^+-c_nc_{n+1}\right),
\;\;\;
c^+_{N+1}=c^+_1,
\;
c_{N+1}=c_1
\end{eqnarray}
(the boundary term, that is unimportant 
as far as the thermodynamics is concerned, has been omitted).
Unlike, in Ref. 6,
we cannot proceed directly  
by using a continued 
fraction representation 
for the diagonal one--fermion Green functions 
of the tight--binding spinless fermions
since the Hamiltonian (\ref{002})
contains the products of two creation
(annihilation) operators.
However, it is well known
(see, e.g., $^{1,11,12}$)
that after introducing new operators
$\eta_k=\sum_{i=1}^N\left(g_{ki}c_i+h_{ki}c_i^+\right)$,
$\eta_k^+=\sum_{i=1}^N\left(h_{ki}c_i+g_{ki}c_i^+\right)$
the Hamiltonian (\ref{002}) transforms into 
\begin{eqnarray}
\label{003}
H=\sum_{k=1}^N\Lambda_k
\left(\eta_k^+\eta_k-\frac{1}{2}\right),
\;\;\;
\left\{\eta_k^+,\eta_q\right\}=\delta_{kq},
\;
\left\{\eta_k,\eta_q\right\}=\left\{\eta_k^+,\eta_q^+\right\}=0
\end{eqnarray}
if the unknown coefficients
$g_{ki}=\frac{1}{2}(\Phi_{ki}+\Psi_{ki})$,
$h_{ki}=\frac{1}{2}(\Phi_{ki}-\Psi_{ki})$
are determined from the following equations
\begin{eqnarray}
\label{004}
\Omega_{n-1}I_{n-1}\Phi_{k,n-1}
+\left(\Omega_n^2+I_{n-1}^2-\Lambda_k^2\right)\Phi_{kn}
+\Omega_nI_n\Phi_{k,n+1}=0,
\;\;\;
\Phi_{k0}=\Phi_{kN},
\;\Phi_{k,N+1}=\Phi_{k1};
\nonumber\\
\Omega_{n}I_{n-1}\Psi_{k,n-1}
+\left(\Omega_n^2+I_{n}^2-\Lambda_k^2\right)\Psi_{kn}
+\Omega_{n+1}I_n\Psi_{k,n+1}=0,
\;\;\;
\Psi_{k0}=\Psi_{kN},
\;\Psi_{k,N+1}=\Psi_{k1}.
\end{eqnarray}
Eqs. (\ref{004}) 
formally coincide with those
describing displacements of particles
in a nonuniform harmonic chain 
with nearest neighbour interactions.
To find the distribution of the squares of ``phonon'' 
(magnetic excitation) frequencies
$R(E^2)=\frac{1}{N}\sum_{k=1}^N\delta(E^2-\Lambda_k^2)$
we may use the Green function approach.
Consider, for example,
the Green functions
$G_{nm}\equiv G_{nm}(E^2)$
that satisfy the set of equations
\begin{eqnarray}
\label{005}
\left(E^2-\Omega_n^2-I_{n-1}^2\right)G_{nm}
-\Omega_{n-1}I_{n-1}G_{n-1,m}
-\Omega_{n}I_{n}G_{n+1,m}
=\delta_{nm}
\end{eqnarray}
with periodic boundary conditions implied.
Knowing $G_{nm}$  
one immediately finds the density of states $R(E^2)$
through the relation
\begin{eqnarray}
\label{006}
R(E^2)=
\mp\frac{1}{\pi N}\sum_{n=1}^N{\mbox{Im}}G_{nn}(E^2\pm i\epsilon),
\;\;\;
\epsilon\rightarrow+0.
\end{eqnarray}
As follows from (\ref{003}) 
all the thermodynamic quantities can be expressed through $R(E^2)$.
For example, the Helmholtz free energy per site is given by
\begin{eqnarray}
\label{007}
f=-\frac{2}{\beta}\int_0^{\infty}{\mbox{d}}E
ER(E^2)\ln\left(2\cosh\frac{\beta E}{2}\right).
\end{eqnarray}
Obviously, $R(E^2)$ can be obtained  
with the help of the Green functions 
introduced on the basis of the set of equations 
for the coefficients $\Psi_{kn}$ (\ref{004}).
We have performed such a calculation
and found the same result for $R(E^2)$
in the cases considered below
(Eqs. (\ref{009}) and (\ref{011})).

We have now to calculate the diagonal Green functions
$G_{nn}$.
Let us use the continued fraction representation
for $G_{nn}$
that follows from (\ref{005})
\begin{eqnarray}
\label{008}
G_{nn}
=\frac{1}
{E^2-\Omega_{n}^2-I_{n-1}^2-\Delta_n^--\Delta_n^+},
\nonumber\\
\Delta_n^-
=\frac{\Omega_{n-1}^2I_{n-1}^2}
{E^2-\Omega_{n-1}^2-I_{n-2}^2
-\frac{\Omega_{n-2}^2I_{n-2}^2}
{E^2-\Omega_{n-2}^2-I_{n-3}^2-_{\ddots}}},
\nonumber\\
\Delta_n^+
=\frac{\Omega_{n}^2I_{n}^2}
{E^2-\Omega_{n+1}^2-I_{n}^2
-\frac{\Omega_{n+1}^2I_{n+1}^2}
{E^2-\Omega_{n+2}^2-I_{n+1}^2-_{\ddots}}}.
\end{eqnarray}
For any finite period of varying
$\Omega_n$ and $I_n$
the continued fractions in (\ref{008}) become periodic
(evidently, the limit $N\to\infty$ is hinted)
and can be easily calculated by solving quadratic equations.
As a result we get rigorous results
for the Green functions,
density of states (\ref{006})
and thermodynamic quantities (\ref{007})
of the periodic alternating spin model (\ref{001}).
Note that the thermodynamic quantities 
are not sensitive to the signs of $\Omega_n$ and $I_n$.
Therefore, one may assume 
$\Omega_n\ge 0$, 
$I_n\ge 0$ 
without loss of generality.

It should be remarked at this point 
that the possibility 
of obtaining $R(E^2)$ exactly 
in the manner described above
exists only for the transverse Ising chain.
For the transverse $XY$ chain with
an arbitrary anisotropic exchange coupling
one arrives at
a set of five diagonal coupled equations
(not three, as above),
corresponding to
a nonuniform harmonic chain 
with nearest 
and next nearest neighbour interactions.$^{1,12}$
Hence, one cannot proceed as above.

To illustrate how the method works
we begin with the uniform case, 
namely, the periodic alternating chain having period 1.
For such a case,
$$\Delta_n^-=\Delta_n^+=\frac{1}{2}(E^2-\Omega^2-I^2
\pm\sqrt{(E^2-\Omega^2-I^2)^2-4\Omega^2I^2}),$$
$$G_{nn}
=\mp\frac{1}{\sqrt{(E^2-\Omega^2-I^2)^2-4\Omega^2I^2}}$$
and therefore
\begin{eqnarray}
\label{009}
R(E^2)=
\left\{
\begin{array}{ll}
\frac{1}{\pi}\frac{1}
{\sqrt{-\left(E^2-a_1\right)
\left(E^2-a_2\right)}}, &
{\mbox{if}}\;\;\;
a_1<E^2<a_2, \\
0, &
{\mbox{otherwise}},
\end{array}
\right.
\end{eqnarray}
with 
$a_1=(\Omega-I)^2$,
$a_2=(\Omega+I)^2$.
From Eq. (\ref{009}),
the ground state energy per site, 
$e_0=-\int_0^{\infty}{\mbox{d}}E E^2 R(E^2)$,
is 
$$e_0=-\frac{1}{\pi}
\int_{\sqrt{a_1}}^{\sqrt{a_2}}
\frac{{\mbox{d}}EE^2}{\sqrt{-(E^2-a_1)(E^2-a_2)}},$$
which after the 
change of variable
$E=\sqrt{a_2-
(a_2-a_1)\sin^2\phi}$,
leads to the known$^{13}$ expression
\begin{eqnarray}
\label{010}
\frac{e_0}{I}
=-\frac{1}{\pi}
\int_{0}^{\frac{\pi}{2}}
{\mbox{d}}\phi\sqrt{\left(1+\lambda\right)^2-4\lambda\sin^2\phi},
\;\;\;
\lambda=\frac{\Omega}{I}.
\end{eqnarray}

Let us now turn to  
the periodic alternating chain having period 2.
Following the procedure above, one finds
\begin{eqnarray}
\label{011}
R(E^2)=
\left\{
\begin{array}{ll}
\frac{1}{2\pi}
\frac{\vert 2E^2-\Omega_1^2-\Omega_2^2-I_1^2-I_2^2\vert}
{\sqrt{{\cal{B}}(E^2)}}, &
{\mbox{if}}\;\;\;
{\cal{B}}(E^2)>0, \\
0, &
{\mbox{otherwise}},
\end{array}
\right.
\nonumber\\
{\cal{B}}(E^2)
=
4\Omega_1^2\Omega_2^2I_1^2I_2^2
-\left(
E^4
-\left(\Omega_1^2+\Omega_2^2+I_1^2+I_2^2\right)E^2
+\Omega_1^2\Omega_2^2+I_1^2I_2^2
\right)^2
\nonumber\\
=-(E^2-b_1)(E^2-b_2)(E^2-b_3)(E^2-b_4),
\nonumber\\
\left\{b_j\right\}
=\left\{
\frac{1}{2}
\left(
\Omega_1^2+\Omega_2^2+I_1^2+I_2^2
\pm\sqrt{(\Omega_1^2+\Omega_2^2+I_1^2+I_2^2)^2
-4(\Omega_1\Omega_2\pm I_1I_2)^2}
\right)
\right\}.
\end{eqnarray}
Obviously, (\ref{011}) 
recovers the result for the uniform chain (\ref{009}) if 
$\Omega_1=\Omega_2=\Omega$,
$I_1=I_2=I$.
The resulting density of states $R(E^2)$ (\ref{011})
yields all thermodynamic quantities 
of the regularly alternating transverse Ising chain 
$\Omega_1I_1\Omega_2I_2\Omega_1I_1\Omega_2I_2\ldots\;$.

Consider further, in some detail,   
the alternating transverse Ising chain with $p=2$ 
with the dimerization ansatz  
$\Omega_1=\Omega_2=\Omega$,
$I_1=I(1+\delta)$,
$I_2=I(1-\delta)$
in mind,
where $0\le\delta\le 1$ is the dimerization parameter.
For such a chain the ground state energy according to (\ref{011})
can be written as 
\begin{eqnarray}
\label{012}
\frac{e_0(\delta)}{I}
=-\frac{1}{2\pi}
\int_0^{\frac{\pi}{2}}
{\mbox{d}}\phi
\left(
\sqrt{1+\lambda^2+\delta^2
+2\sqrt{W} }
+\sqrt{1+\lambda^2+\delta^2
-2\sqrt{W} }
\right),
\nonumber\\
W=\lambda^2+\delta^2-\lambda^2(1-\delta^2)\sin^2\phi.
\end{eqnarray}
If $\delta=0$ (\ref{012}) 
transforms into the ground state energy of the uniform chain (\ref{010}).

Further,
we calculate the zero temperature transverse magnetization 
$m_z=\frac{\partial e_0}{\partial\Omega}$ 
by differentiating the r.h.s of Eq. (\ref{012}) 
with respect to $\lambda$. 
In Fig. 1 the obtained dependence 
of the transverse magnetization on the transverse field  
for several values of the dimerization parameter $\delta$
is plotted. 
As one can see from Fig. 1 
the transverse Ising chain with regularly alternating bonds 
does not exhibit 
a step--like dependence of magnetization vs. field
and the magnetization profile only smoothly deforms 
with the increasing $\delta$.
This behaviour is contrary to that of the
transverse $XY$ chain
(for which the Hamiltonian (\ref{001}) 
would contain
$s^x_ns^x_{n+1}+s^y_ns^y_{n+1}$ instead of $s^x_ns^x_{n+1}$), 
since the latter model  
does show a step--like dependence 
of magnetization vs. field
(i.e. a plateau at $m_z=0$)
for regularly alternating bonds$^{6}$
(the light curves in Fig. 1).
The dissimilarity is conditioned 
by the different symmetries of these models,
since 
$\sum_{n=1}^Ns_n^z$
does not commute with the Hamiltonian of the transverse Ising chain 
but it does commute with the Hamiltonian of the transverse $XY$ chain.
It is worth to consider in addition both models in the limit 
$\delta=1$ 
when the chain splits into noninteracting clusters
consisting of two sites.
The zero temperature transverse magnetization follows from the formula 
$m_z=\frac{1}{2}\langle GS\vert s_1^z+s_2^z\vert GS \rangle$ 
where 
$\vert GS \rangle$ 
is the ground state eigenvector 
of the two--site cluster Hamiltonian. 
For the transverse Ising model 
$\vert GS \rangle$ 
smoothly varies with increasing $\Omega$.
Moreover,
$\langle GS\vert s_1^z+s_2^z\vert GS \rangle=0$
if $\Omega=0$
and 
$\langle GS\vert s_1^z+s_2^z\vert GS \rangle\to -1$
if $\Omega\to \infty$.
Unlike,
for the transverse $XY$ model 
the ground state is singlet 
(and thus $\langle GS\vert s_1^z+s_2^z\vert GS \rangle =0$)
if $0\le\Omega<2 I$.
However, 
if $\Omega$ exceeds $2I$ 
the ground state becomes triplet 
and $\langle GS\vert s_1^z+s_2^z\vert GS \rangle$
abruptly changes to $-1$.

The ground state static transverse susceptibility 
$\chi_{zz}
=\frac{\partial m_z}{\partial\Omega}
=\frac{1}{I}\frac{\partial^2}{\partial\lambda^2}\frac{e_0(\delta)}{I}$
follows straightforwardly from Eq. (\ref{012}).
In the left panel in Fig. 2 
the obtained dependence 
of the static transverse susceptibility 
on the transverse field 
for several $\delta$ is plotted.
Using Eq. (\ref{012}) it is easy to show 
that $\chi_{zz}$ for the dimerized chain consists of 
a finite part and a singular part
\begin{eqnarray}
\label{013}
\chi_{zz}
={\mbox{finite term}}
-\frac{\lambda^2}{2\pi}
\int_0^{\frac{\pi}{2}}{\mbox{d}}\phi
\frac{\left(1-\left(1-\delta^2\right)\sin^2\phi\right)^2}
{W^{\frac{3}{2}}
\sqrt{1+\lambda^2+\delta^2-2\sqrt{W}}}.
\end{eqnarray}
As it follows from (\ref{013}) 
the ground state static transverse susceptibility 
exhibits a logarithmic singularity 
at $\lambda=\lambda^{\star}=\sqrt{1-\delta^2}$,
i.e. 
$\chi_{zz}\sim \ln\vert\lambda-\lambda^{\star}\vert$,
$\lambda\to\lambda^{\star}$. 
It should be noted, 
that $\chi_{zz}$ for the corresponding dimerized $XY$ chain 
exhibits two square root singularities
at $\lambda^{\star}_1=2\delta$ and $\lambda^{\star}_2=2$
(right panel in Fig. 2, see also Ref. 6).
 
To discuss a spin--Peierls dimerization 
in the adiabatic limit 
one must examine the dependence of the total energy
${\cal{E}}(\delta)=e_0(\delta)+\alpha\delta^2$, 
which consists of the magnetic part $e_0(\delta)$ (\ref{012}) 
and the elastic part $\alpha\delta^2$, $\alpha>0$, 
on the dimerization parameter $\delta$.
In Fig. 3 
one can see  
the dependence of the magnetic energy (\ref{012}) on $\delta$ 
for different values of the transverse field.
From these plots one concludes
that the magnetic energy 
as a rule 
decreases with the increase of $\delta$
(except for $\lambda=0$ and $\lambda\to\infty$ 
when $\frac{e_0(\delta)}{I}$ (\ref{012}) 
does not depend on $\delta$ 
and equals $-\frac{1}{2}$ and $-\frac{\lambda}{2}$, respectively) 
that is a necessary condition 
for the existence of the dimerized phase.
However, the magnetic energy decreases too slowly
in comparison with the increase of the elastic energy 
that provides stability of the chain with respect to dimerization.
To demonstrate this, 
we calculate 
$a=-\frac{\partial}{\partial\delta^2}\frac{e_0(\delta)}{I}$ 
by differentiating 
the r.h.s. of Eq. (\ref{012}) with respect to $\delta^2$, 
and display the dependence of this quantity on $\delta$  
for several values of the transverse field 
in Fig. 4. 
These plots show  
that ${\cal{E}}(\delta)$ may exhibit a maximum 
(but not a minimum) 
at the nonzero value of dimerization parameter 
for lattices having small $\frac{\alpha}{I}$
as the slope of the curves 
$a=-\frac{\partial}{\partial\delta^2}\frac{e_0(\delta)}{I}$
vs. $\delta$ 
manifests. 
Hence, the uniform chain with $\delta=0$ is favourable.
On the contrary, for the $XY$ chain 
the ground state energy 
decreases sufficiently rapidly to 
provide a stability of the dimerized phase.
In Fig. 4 we also display the dependence of 
$a=-\frac{\partial}{\partial\delta^2}\frac{e_0(\delta)}{I}$ 
on $\delta$ for the transverse $XY$ chain 
(light curves)  
(see also Ref. 6)
emphasizing explicitly the difference 
between those two spin models.

It is worthy to note, 
that the absence of a spin--Peierls dimerization 
in the transverse Ising chain may be anticipated 
on the basis of the following reasoning.$^{14}$
Considering the transverse $XY$ chain 
one notes that for $\Omega=0$
the ground state corresponds to a half--filled fermion band,
the system is gapless 
and the Peierls mechanism does work.
As $\Omega$ increasing becomes equal to $I$
the ground state corresponds to an empty fermion band,
the system becomes gaped
and the dimerized phase does not appear.
The ground state of the transverse Ising chain$^{13}$ 
always corresponds to an empty fermion band,
the system is gapped unless $\Omega=I$ 
and as a result no Peierls dimerization should be expected.

To conclude,
we have proposed to use continued fractions 
for the calculation of thermodynamic quantities 
of regularly alternating spin--$\frac{1}{2}$ transverse Ising chains. 
For such a chain, 
in which the bonds 
are alternating in magnitude along the chain with period 2, 
we have found no plateaus in the dependence 
magnetization vs. field. 
The susceptibility of the chain exhibits a logarithmic singularity 
at the field determined by interaction strengths.
Moreover, 
we have found that in the adiabatic limit 
the transverse Ising chain  
is stable with respect to a spin--Peierls dimerization. 
These findings differ dramatically  
from the corresponding results  
for the transverse isotropic $XY$ chain, 
and therefore 
it would be of great interest to study 
the periodic alternating transverse {\em {anisotropic}} $XY$ chain 
that connects both these limiting cases.

The author is grateful for the hospitality 
to the Max Planck Institute for the Physics of Complex Systems, Dresden 
at the end of 1999 
when the paper was completed. 
The paper was presented in part 
at the International Workshop on the Theme of 
``Microscopic Theories of Phase Transitions: 
Quantum Versus Thermal Fluctuations" 
(Bad Honnef, 1999). 
The author thanks the WE--Heraeus Foundation for the hospitality  
and the participants for discussions and comments.
He acknowledges numerous valuable discussions with 
J. Richter, T. Krokhmalskii and O. Zaburannyi
and expresses appreciation to J. W. Tucker 
for a critical reading of the manuscript.
Finally, he thanks the referees 
for many useful comments and helpful suggestions.

\vspace{5mm}

{\large{\bf List of figure captions}}
\normalsize

\vspace{1mm}

FIGURE 1. 
The ground state transverse magnetization vs. transverse field 
for the transverse Ising chain with regularly 
alternating bonds with the period 2.
$\delta=0$ (uniform chain),
$\delta=0.3$,
$\delta=0.6$,
and 
$\delta=0.9$
correspond to 
solid, long--dashed, short--dashed, and dotted  
bold curves, respectively. 
For comparison the corresponding results for the transverse $XY$ chain 
are shown by light curves.   

\vspace{1mm}

FIGURE 2.
The ground state static transverse susceptibility vs. transverse field 
for the transverse Ising chain (left panel)
and the transverse $XY$ chain (right panel)
with regularly alternating bonds with the period 2.
$\delta=0$ (uniform chain),
$\delta=0.3$,
$\delta=0.6$,
and 
$\delta=0.9$
correspond to 
solid, long--dashed, short--dashed, and dotted  
curves, respectively. 

\vspace{1mm}

FIGURE 3.
The change of the ground state magnetic energy  
$d=\frac{e_0(\delta)}{I}-\frac{e_0(0)}{I}$
vs. $\delta$ 
for the transverse Ising chain 
for $\lambda=0,\;0.25,\;0.5,\;1,$ and $2$ 
(solid, long--dashed, short--dashed, dotted, and dashed--dotted   
bold curves, 
respectively).
For comparison the corresponding results for the transverse $XY$ chain 
are shown by light curves.   

\vspace{1mm}

FIGURE 4.
The dependence 
$a=-\frac{\partial}{\partial\delta^2}\frac{e_0(\delta)}{I}$ 
vs. $\delta$
for the transverse Ising chain 
with $\lambda=0,\;0.25,\;0.5,\;1,$ and $2$ 
(solid, long--dashed, short--dashed, dotted, and dashed--dotted   
bold curves, respectively).
For comparison the corresponding results for the transverse $XY$ chain 
are shown by light curves.   

\clearpage

\begin{figure}
\vspace{0mm}
\epsfxsize=100mm
\epsfbox{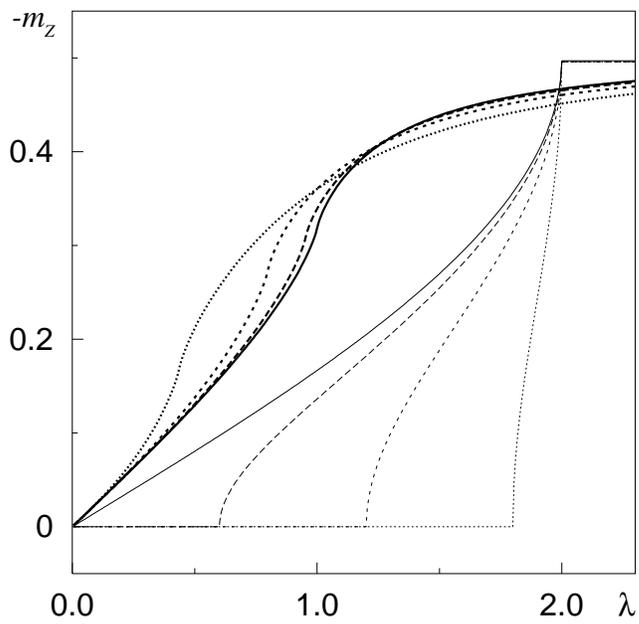}
\vspace{0mm}
\caption{\small{
The ground state transverse magnetization vs. transverse field 
for the transverse Ising chain with regularly 
alternating bonds with the period 2.
$\delta=0$ (uniform chain),
$\delta=0.3$,
$\delta=0.6$,
and 
$\delta=0.9$
correspond to 
solid, long--dashed, short--dashed, and dotted  
bold curves, respectively. 
For comparison the corresponding results for the transverse $XY$ chain 
are shown by light curves.   
}}
\end{figure}

\clearpage

\begin{figure}
\vspace{0mm}
\epsfxsize=150mm
\epsfbox{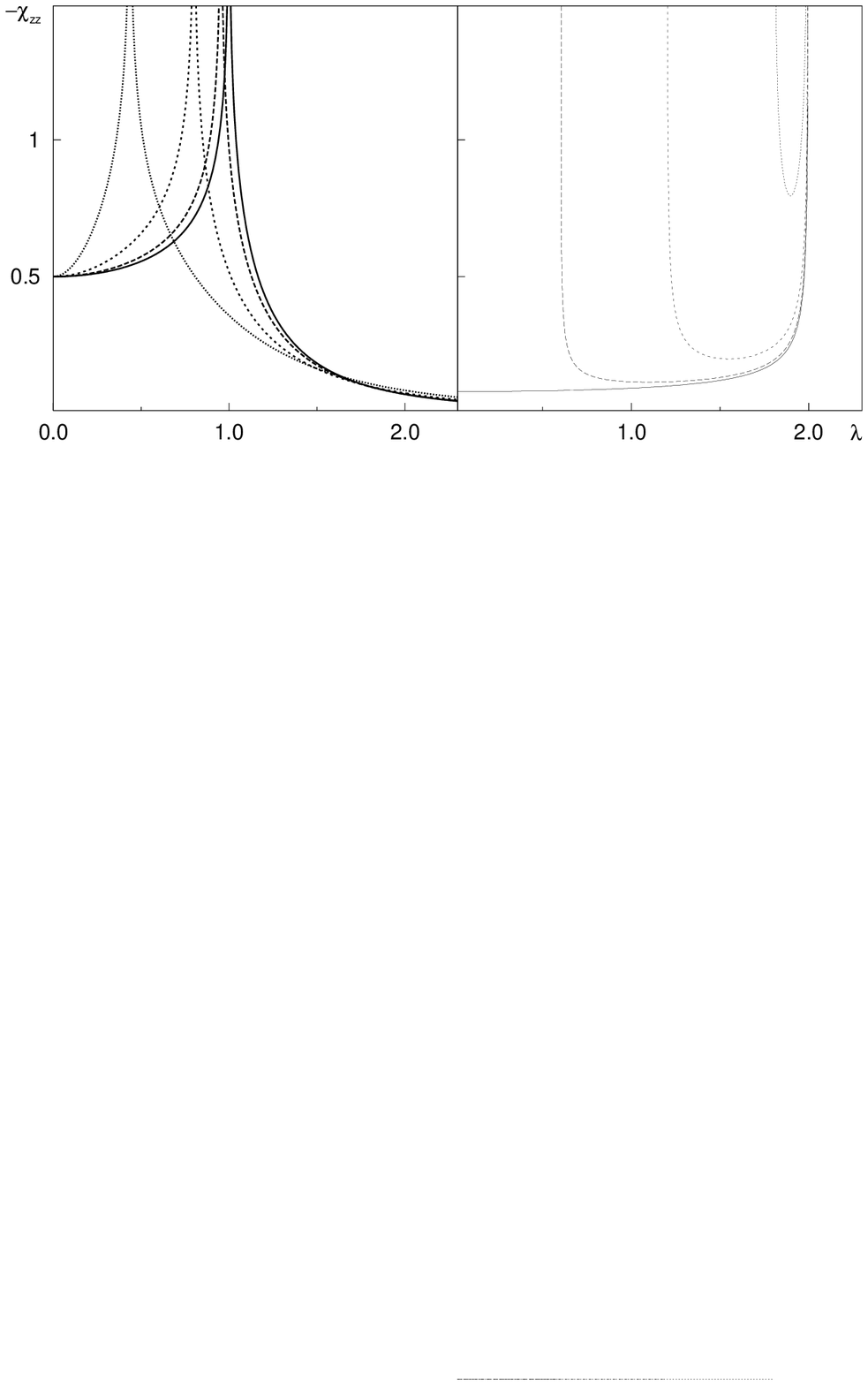}
\vspace{0mm}
\caption{\small{
The ground state static transverse susceptibility vs. transverse field 
for the transverse Ising chain (left panel)
and the transverse $XY$ chain (right panel)
with regularly alternating bonds with the period 2.
$\delta=0$ (uniform chain),
$\delta=0.3$,
$\delta=0.6$,
and 
$\delta=0.9$
correspond to 
solid, long--dashed, short--dashed, and dotted  
curves, respectively. 
}}
\end{figure}

\clearpage

\begin{figure}
\vspace{0mm}
\epsfxsize=100mm
\epsfbox{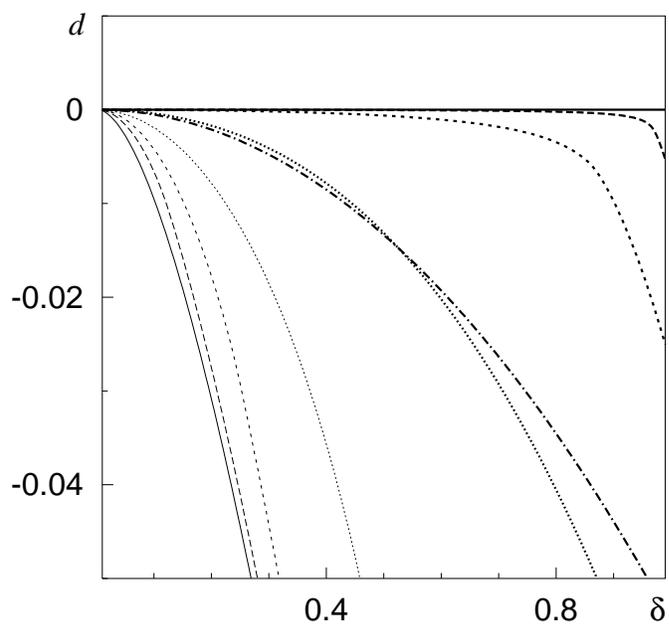}
\vspace{0mm}
\caption{\small{
The change of the ground state magnetic energy  
$d=\frac{e_0(\delta)}{I}-\frac{e_0(0)}{I}$
vs. $\delta$ 
for the transverse Ising chain 
for $\lambda=0,\;0.25,\;0.5,\;1,$ and $2$ 
(solid, long--dashed, short--dashed, dotted, and dashed--dotted   
bold curves, 
respectively).
For comparison the corresponding results for the transverse $XY$ chain 
are shown by light curves.   
}}
\end{figure}

\clearpage

\begin{figure}
\vspace{0mm}
\epsfxsize=100mm
\epsfbox{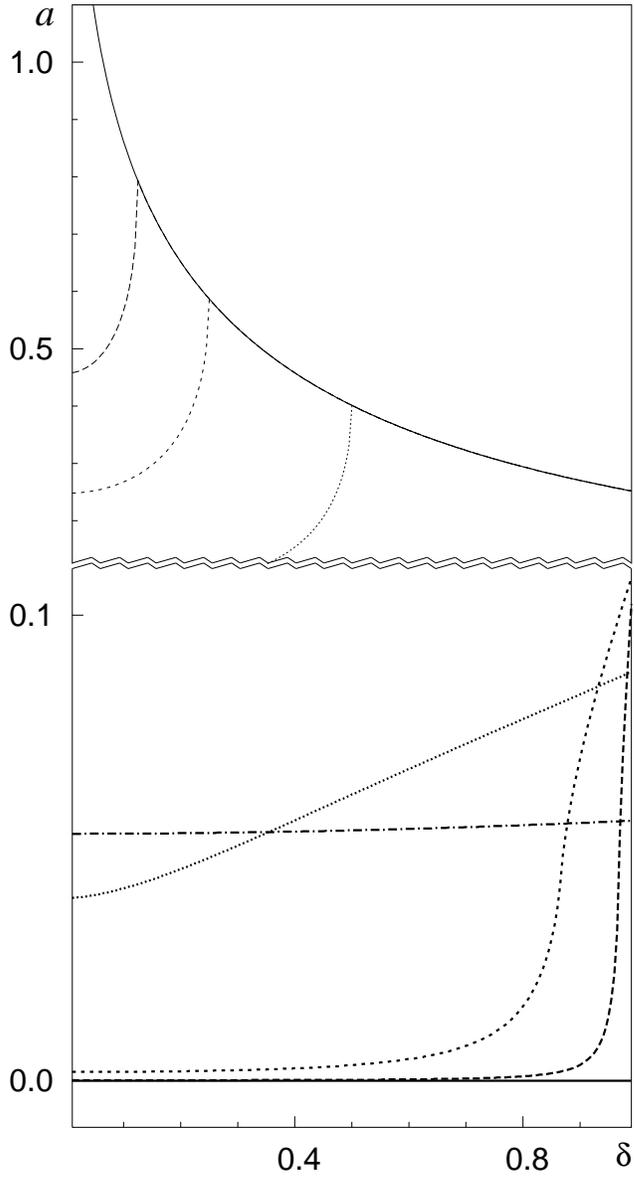}
\vspace{0mm}
\caption{\small{
The dependence 
$a=-\frac{\partial}{\partial\delta^2}\frac{e_0(\delta)}{I}$ 
vs. $\delta$
for the transverse Ising chain 
with $\lambda=0,\;0.25,\;0.5,\;1,$ and $2$ 
(solid, long--dashed, short--dashed, dotted, and dashed--dotted   
bold curves, respectively).
For comparison the corresponding results for the transverse $XY$ chain 
are shown by light curves.   
}}
\end{figure}


\begin{thebibliography}{}
\bibitem{014}
These arguments were suggested by an anonymous referee.
\end{thebibliography}
\end{document}